# High Curie temperature and perpendicular magnetic anisotropy in homoepitaxial InMnAs films

Y Yuan[1, 4], Y Wang[1,4], K Gao[1, 4], M Khalid[1], C Wu[1, 2], W Zhang[2], F Munnik[1], E Weschke[3], C Baehtz[1], W Skorupa[1], M Helm[1, 4], and S Zhou[1]

[1]Helmholtz-Zentrum Dresden Rossendorf, Institute of Ion Beam Physics and Materials Research, Bautzner Landstrasse 400, D-01328 Dresden, Germany

[2]State Key Lab of Electronic Thin films and Integrated Devices, University of Electronic Science and Technology of China, Chengdu 610054, China

[3]Helmholtz-Zentrum Berlin für Materialien und Energie, Wilhelm-Conrad-Röntgen-Campus BESSY II, D-12489 Berlin, Germany

[4]Technische Universität Dresden, D-01062 Dresden, Germany

**Abstract:**

We have prepared the dilute magnetic semiconductor (DMS) InMnAs with different Mn concentrations by ion implantation and pulsed laser melting. The Curie temperature of the $In_{1-x}Mn_xAs$ epilayer depends on the Mn concentration x, reaching 82 K for x=0.105. The substitution of Mn ions at the Indium sites induces a compressive strain perpendicular to the InMnAs layer and a tensile strain along the in-plane direction. This gives rise to a large perpendicular magnetic anisotropy, which is often needed for the demonstration of electrical control of magnetization and for spin-transfer-torque induced magnetization reversal.

Studies focusing on the preparation of III-V dilute magnetic semiconductors (DMS) exhibiting a high Curie temperature ($T_C$) have been greatly inspired due to potential applications in spintronic devices [1-5]. The carrier mediated nature of the ferromagnetism in DMS allows for controlling the magnetic properties (Curie temperature, anisotropy and etc.) by an electric field [6, 7]. Current induced domain wall switching and motion have been demonstrated in GaMnAs [4, 5, 8] and GaMnAsP quaternary alloy [9-11]. In those investigations, a perpendicular magnetic anisotropy (PMA) is required for the ferromagnetic semiconductors [8-11]. Homoepitaxial GaMnAs on GaAs substrate normally exhibits in-plane magnetic anisotropy due to the compressive strain along its in-plane direction. The PMA in GaMnAs has been realized by two approaches: The mostly used one is to tune the strain from compressive to tensile state in GaMnAs either by applying InGaAs as buffer [8, 9] or by co-alloying GaMnAs with phosphorus [10-13]. The other one is to change the localization of holes in the GaMnAs layer, by co-alloying Al [14-17]. Different from GaMnAs, homoepitaxial InMnAs on InAs is subject to tensile strain along its in-plane direction, therefore, should exhibit PMA. Moreover, ferromagnetic InMnAs films in principle provide a more suitable test-bed for current induced magnetic switching in magnetic semiconductors due to its larger spin-orbit coupling [18].

The preparation of InMnAs by low-temperature molecular beam epitaxy (LT-MBE) was not satisfactory in the early stage: the large lattice mismatch between the layer and the substrate results in a large amount of n-type defects, which are detrimental to magnetization and suppress $T_C$ [19, 20]. Schallenberg and Munekata prepared recently InMnAs with the highest $T_C$ of 90 K through LT-MBE by carefully selecting a buffer layer [21]. However, once the strain is completely relaxed, the PMA is lost in the InMnAs layer. So far, the highest $T_C$ of InMnAs films with PMA is around 72 K when the Mn concentration reaches to 10% grown on an AlSb buffer by LT-MBE [21], i. e. still below the liquid nitrogen temperature. Thus, the preparation of InMnAs with PMA and a higher $T_C$ (above the liquid nitrogen temperature in the view of technical practice) is still challenging and interesting especially when it is prepared through a simple and versatile method.

In this letter, we present the preparation of homoepitaxial InMnAs on InAs by ion implantation and pulsed laser melting. We show that a highest $T_C$ of 82 K is achieved in $In_{1-x}Mn_xAs$ with PMA at a Mn concentration of x=0.105. The magnetic anisotropy in InMnAs is confirmed by magnetization and x-ray magnetic circular dichroism (XMCD) measurements. The pronounced PMA in this system is attributed to the tensile strain in the in-plane direction of InMnAs arising from the Mn ion incorporation.

InMnAs epilayers with different Mn concentrations were prepared by Mn ion implantation and pulsed laser melting on InAs (001) substrates. First, Mn ions were implanted into intrinsic InAs substrates with an energy of 100 keV. According to SRIM simulation [22], the projected range ($R_P$) and the longitudinal straggling ($\Delta R_P$) for Mn distribution are around 60 and 38 nm, respectively. The Mn fluences ($\Phi$) were $4\times10^{15}$, $8\times10^{15}$, $1.6\times10^{16}$, $2.4\times10^{16}$, and $2.8\times10^{16}$ cm$^{-2}$. A XeCl excimer laser (Coherent ComPexPRO201, wavelength of 308 nm and pulse length of 30 ns) with an energy density of 0.30 J/cm$^2$ was used for annealing the InMnAs samples. After the pulsed laser annealing and HCl etching, the remaining Mn fluences determined by particle induced X-ray emission (PIXE) using 3 MeV protons are $3.25\times10^{15}$, $7.26\times10^{15}$, $1.18\times10^{16}$, $1.48\times10^{16}$, and $1.78\times10^{16}$ cm$^{-2}$, respectively. The virtual Mn concentration can be calculated by $\Phi/\sqrt{2\pi}\Delta R_p$. According to the remaining fluence, the Mn concentration x ($In_{1-x}Mn_xAs$) is determined to be 0.019, 0.042, 0.069, 0.087, and 0.105, respectively. Magnetic properties were measured by a Superconducting Quantum Interference Device (SQUID) magnetometer (Quantum Design, SQUID-VSM). Synchrotron radiation x-ray diffraction (SRXRD) was performed on the InMnAs samples at the BM20 (ROBL) beamline at the European Synchrotron Radiation Facility (ESRF). It was using a wavelength of λ=1.0781 Å. X-ray absorption spectroscopy (XAS) measurements were performed at the beamline UE46 / PGM-1 at BESSY II (Helmholtz-Zentrum Berlin). A magnetic field up to 1 T was applied parallel or anti-parallel to the photon helicity and the samples were rotated from perpendicular to the field to different angles. Before XAS measurements, all the samples were dipped in HCl solution to remove the native oxide layer from the surface.

Figure 1(a) represents the results of magnetic field dependent magnetization measured at 5 K for InMnAs with different Mn concentrations. Note that the diamagnetic background is subtracted from the data shown in Fig. 1(a). All samples were annealed with an energy density of 0.30 J/cm$^2$. The magnetic field is applied perpendicular to the layer (i.e. the field is parallel to the InMnAs [001] axis). At a magnetic field of 300 Oe, all the samples attain their saturation magnetization, and exhibit square-like hysteresis loops which indicate that the out-of-plane direction is the magnetic easy axis for all samples. The magnetic remanence as a function of temperature is plotted in Figure 1(b) for three samples. The samples were cooled down from room temperature under a magnetic field of 1000 Oe. Then the field was decreased to 20 Oe, which should compensate the possible residual field of the superconducting magnet. The magnetic remanence (Mr) was measured during warming up. From the remanence measurements, $T_C$ around 14, 40, and 77 K for x=0.019, 0.042, and 0.069 for $In_{1-x}Mn_xAs$ samples, respectively, are obtained. The inset to Figure 1(b) shows the behavior of Mr(T)/Mr(5K) as a function of T/$T_C$, and the curves follow the mean-field-like approximation [23]. This also indicates that a lot of Mn ions replace Indium ions resulting in a large hole concentration which eventually induces carrier mediated ferromagnetism in InMnAs.

We increased the Mn implantation fluence to check if we can obtain higher Curie temperature. Figure 2(a) shows the relationship between the remaining Mn fluence determined by PIXE and the nominal implantation fluence. At the low fluence regime, most of implanted Mn ions are remaining in the InAs matrix, however, the remaining fraction is drastically decreased with increasing implantation fluence. This large reduction is mainly due to the partitioning at the solid/liquid interface during pulsed laser annealing [24]. In the case of Mn implanted GaAs and GaP, it was found that around 50% Mn can be rejected [24]. Moreover, at large ion fluence the sputtering effect arises and sets the upper limit for the Mn concentration to be implanted [25]. The Curie temperature and magnetization of $In_{1-x}Mn_xAs$ samples as a function of Mn concentration x are displayed in Figure 2(b). The Curie temperature monotonically increases from 14 to 77 K when x is from 0.019 to 0.069. However, the Curie temperature seems to saturate at around 82 K with further increasing Mn concentration up to 0.105. So far, the highest $T_C$ in InMnAs with PMA is achieved in our sample, i.e., 82 K, which is accessible by liquid nitrogen cooling. Meanwhile, the same saturation tendency is seen for the magnetization at 5 K shown in Figure 2(b).

We also calculate the saturation magnetization per Mn by assuming that all remaining Mn ions substitute on the Indium sites. Except for the sample with x of 0.019, the saturation magnetization is calculated to be between 2.3 and 2.5 $\mu_B$/Mn. It is smaller than the theoretical value for GaMnAs (~ 4 $\mu_B$/Mn) [26], however comparable with the experimental values (~ 2 $\mu_B$/Mn in Ref. 26 and ~ 3 $\mu_B$/Mn in Ref. 27) for the state-of-art GaMnAs films. In reality, the substitutional concentration of Mn should be smaller than what we determined by PIXE and there are always defects (for instance As antisites [26]) compensating the holes. Both facts lead the underestimation for the magnetization in term of $\mu_B$/Mn. Indeed, the saturation magnetization is comparable to those obtained for GaMnP (2.8 $\mu_B$/Mn) and GaMnAs (3.4 $\mu_B$/Mn) epilayers prepared by ion implantation [28-30].

Figure 3 shows the magnetic hysteresis loops (normalized) of $In_{0.931}Mn_{0.069}As$ measured at 5 K in two configurations e.g., when the field is applied parallel to the [001] direction (out-of-plane) or [1-10] direction (in-plane). A clear square-like hysteresis loop is observed when the magnetic field is applied along InMnAs [001] direction. It saturates at a low magnetic field compared to another field configuration (parallel to the sample plane). This again proves that the $In_{0.931}Mn_{0.069}As$ sample has an out-of-plane magnetic easy axis. By using the equation $H_d = 4\pi M_S$, where $M_S$ is the saturation magnetization (26 emu/cm$^3$ for $In_{0.931}Mn_{0.069}As$), we can calculate the demagnetizing field induced by the shape anisotropy to be 0.32 kOe. The lowest-order uniaxial magnetic anisotropy field induced by strain is 2.16 kOe obtained by the equation $H_{u\perp} = [M_S/(dM/dH)_{H=0}] + H_d$ [17]. Different from the homoepitaxial GaMnAs on GaAs synthesized by LT-MBE, our samples do not show any temperature dependent magnetic anisotropy [31]. Note that the substitution of Indium by Mn results in a smaller lattice constant and compressive strain along the InMnAs [001] axis, and this strain is mainly responsible for the uniaxial perpendicular magnetic anisotropy [32]. Due to the biaxial compressive strain, the valence band splits and the lowest valence band assumes a heavy-hole character. The hole spins are oriented along the growth direction when only the lowest valence band is occupied, since in this case it can lower their energy by coupling to the Mn spins and PMA is expected [31].

The most probable secondary phase is MnAs, like in the case of Mn implanted GaAs upon rapid thermal annealing [33, 34]. For bulk MnAs or nanocrystalline MnAs in GaAs, the Curie temperature is above 300 K [33, 34]. As shown in Figure 1(b) and Figure 3, there is no indication of secondary magnetic phase with a higher $T_C$. Moreover, nanocrystalline MnAs embedded in GaAs reveals an easy axis along the in-plane of the film [33, 34]. Therefore, the ferromagnetism in the InMnAs film is intrinsic and the perpendicular magnetic anisotropy is due to the strain. The existence of strain is proved by XRD (see Figure 4) and is also in agreement with the mean-field theory for dilute magnetic semiconductors [32].

The crystalline structure of the homoepitaxial InMnAs is investigated using XRD. The θ-2θ scan around the InAs (004) reflection in Figure 4(a) shows a broad InMnAs peak located at a higher angle than the InAs (004) peak. It means that the InMnAs layer is under a compressive strain along the out-of-plane direction. The presence of x-ray Pendellösung fringes around InAs (004) proves a good crystalline quality of the InMnAs layer. In order to get more detailed information about the structure of InMnAs, a simulation is performed to fit the experimental XRD data of $In_{0.931}Mn_{0.069}As$ as shown in Figure 4(a). The fitting reveals a compressive strain $(a-a_0)/a_0$ of -0.19% for the $In_{0.931}Mn_{0.069}As$ layer.

Figure 4(b) shows the reciprocal space map (RSM) for the (422) reflection from the sample $In_{0.931}Mn_{0.069}As$. Similar to the (004) reflection, a characteristic InMnAs peak appears at a larger $q_z$ in the (422) reflection. The peak is located at the same $q_x$ as the InAs (442) peak. This shows that the InMnAs layer is fully strained on the InAs substrate. The XRD results prove that the InMnAs layer on InAs substrate exhibits a compressive strain along the perpendicular direction and the strain is not relaxed.

Figure 5(a) shows the XAS results of $In_{0.931}Mn_{0.069}As$ at the Mn $L_{2,3}$ edges measured in the total electron yield (TEY) mode under a magnetic field of 1 T at 4.2 K. The features around 640 eV and 651 eV correspond to the $L_3$ ($2p_{3/2}$ to 3d) and $L_2$ ($2p_{1/2}$ to 3d) levels, similar to the situation of GaMnAs [35]. We did not observe any pronounced multiplet structure around the $L_3$ edge which indicates that the HCl etching has effectively removed the oxide layer from the InMnAs surface. The XMCD was already estimated in our earlier work, where the maximum asymmetry was around 51 % at the $L_3$ edge [36]. Tensile strain exists in the in-plane direction and compressive in the out-of-plane direction, which leads to a large uniaxial anisotropy. This uniaxial magnetic anisotropy is related to the special feature in the pre-edge region at 0.5 eV below the $L_3$ peak in the XMCD spectrum [16, 37, 38]. Figure 5(b) presents the angular dependence of the XMCD spectrum of $In_{0.931}Mn_{0.069}As$, where the angular θ is varied from 0° (normal incidence) to 70° (grazing incidence). When rotating the sample from normal to grazing angle, the height of the pre-edge shoulder decreases, which means the compressive strain transforms to tensile strain from the out-of-plane to the in-plane direction. This behavior is similar to other DMS materials, e.g. GaMnAs and GaMnAsP [37, 38], which have a compressive strain along the perpendicular or parallel direction. Figure 5(c) displays the XMCD spectra (normalized at the Mn $L_3$ peak) for $In_{1-x}Mn_xAs$ samples with x=0.042 and 0.069 at a normal incident angular. An increase in the pre-edge shoulder indicates that more Mn ions are incorporated into the InMnAs lattice which results in a larger compressive strain along the perpendicular direction.

Another evidence to prove the carrier-mediated nature of the ferromagnetism in InMnAs layers is magneto-transport. However, due to the narrow bandgap of InAs, the substrate is highly conductive, preventing magneto-transport measurements for the InMnAs layer. The conducting InAs substrate with a large carrier mobility also results in strong Shubnikov-de Haas oscillations in the ferromagnetic resonance (FMR, not shown), over-dominating the features of the InMnAs layer. Otherwise, the Gilbert damping coefficient can be obtained by FMR. The Gilbert damping coefficient is an important parameter for spintronics and has been determined to be in the range of 0.12-0.24 for GaMnAs [39, 40].

In summary, we have prepared ferromagnetic InMnAs with a perpendicular magnetic anisotropy and a high Curie temperature up to 82 K. The XMCD results also confirm the existence of a perpendicular magnetic anisotropy in InMnAs epilayers. To our knowledge, this is the highest $T_C$ recorded for InMnAs with perpendicular magnetic anisotropy. The good crystalline quality of these epilayers is confirmed by the appearance of the Pendellösung fringes around the InAs (004) peak in the XRD pattern and by the square-like hysteresis loops along the magnetic easy axis. The inner strain induced by the Mn ion substitution in the InMnAs epilayers is responsible for the appearance of a perpendicular magnetic anisotropy.

This work is funded by the Helmholtz-Gemeinschaft Deutscher Forschungszentren (HGF-VH-NG-713). Ion implantation was carried out in the Ion Beam Center at HZDR. The author Y. Y. (File No.201306120027) and Y. W. (File No. 2010675001) thank the financial support by China Scholarship Council. C. W. and W. Z. acknowledge the support by the International Science and Technology Cooperation Program of China (2012DFA51430).

**Figure captions**

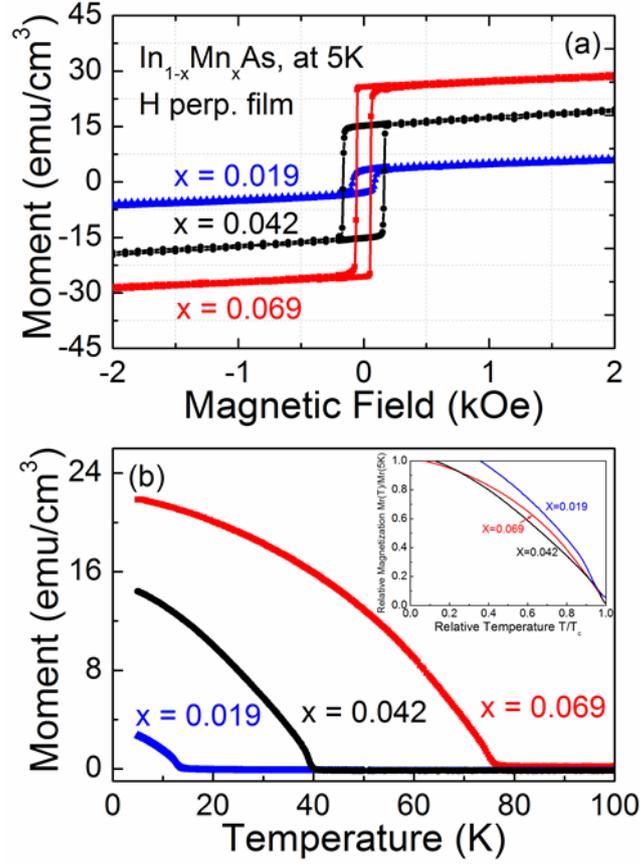

FIG. 1. (Color online) (a) Magnetization of $In_{1-x}Mn_xAs$ samples measured at 5 K in the out-of-plane geometry: The Mn concentrations are: x=0.019 (triangle), 0.042 (circle), and 0.069 (square); (b) Remanent magnetization as a function of temperature measured under a magnetic field of 20 Oe after the samples were saturated by applying 1000 Oe along the perpendicular direction for the three samples. The inset shows the plots of Mr(T)/Mr(5K) as a function of $T/T_C$.

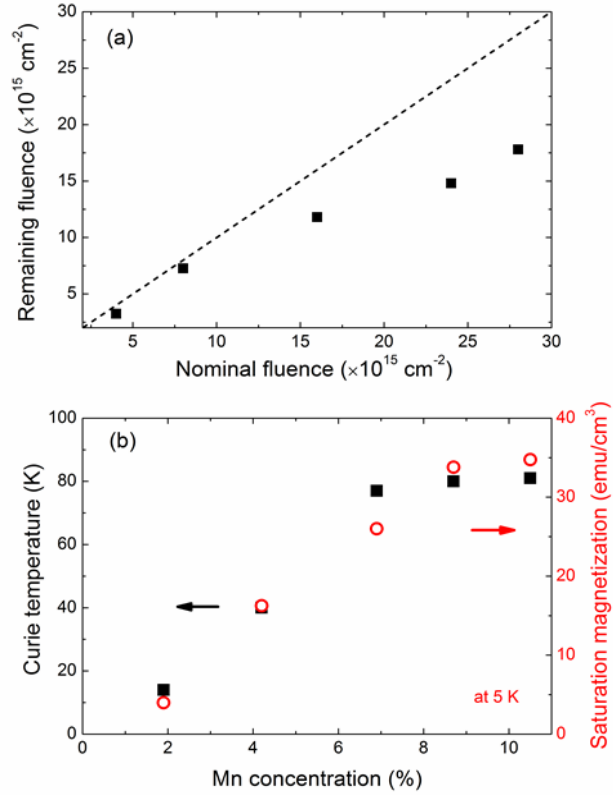

FIG. 2. (Color online) (a) The remaining Mn fluence determined by PIXE vs. the nominal Mn implantation fluence. The dashed line shows the 1:1 relation. (b) Curie temperature and magnetization of $In_{1-x}Mn_xAs$ samples measured as a function of Mn concentration x. The five samples in (b) are the same as shown in (a).

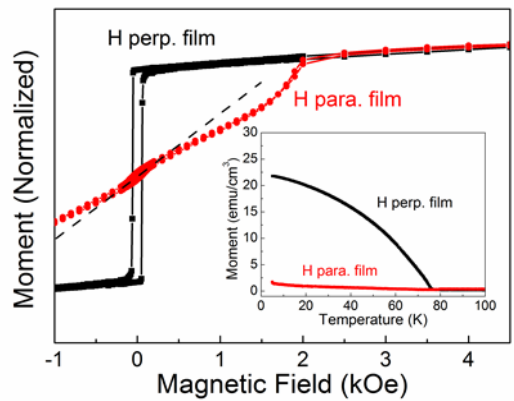

FIG. 3. (Color online) Magnetic field dependence of magnetization of $In_{0.931}Mn_{0.069}As$ with the field perpendicular to the surface (along InMnAs [001]): square and parallel to the surface: circle, at 5 K. The inset shows the remanent magnetization vs. temperature for the different orientations.

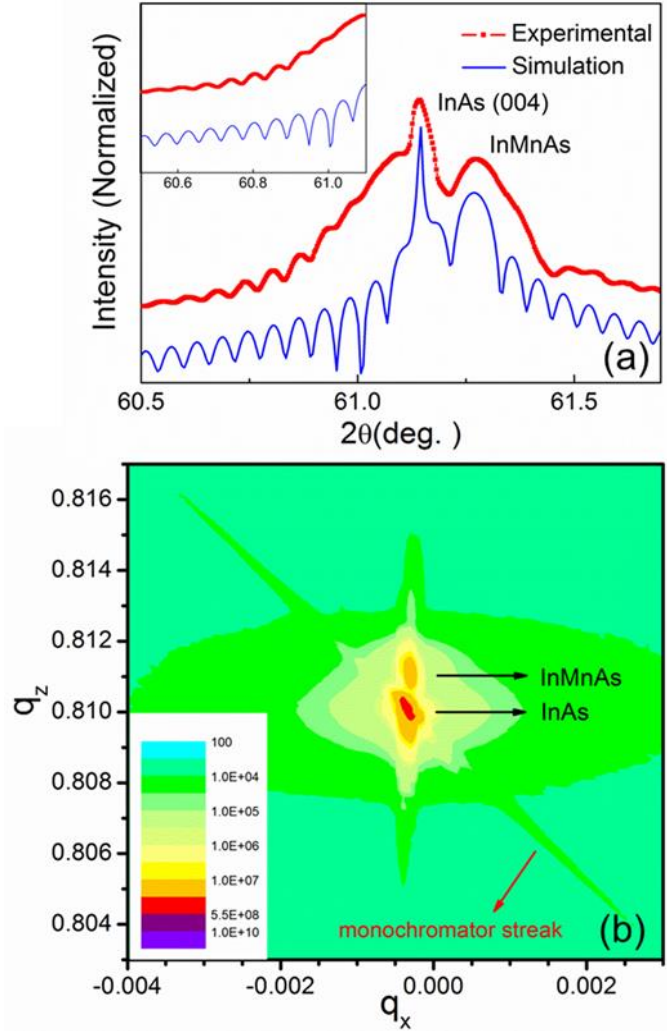

FIG. 4. (Color online) (a) XRD θ-2θ scan (square) and simulation (line) of the (004) reflection for $In_{0.931}Mn_{0.069}As$ sample (the data have been transformed to λ=1.54056 Å); (b) Reciprocal space map (RSM) around the (422) reflection.

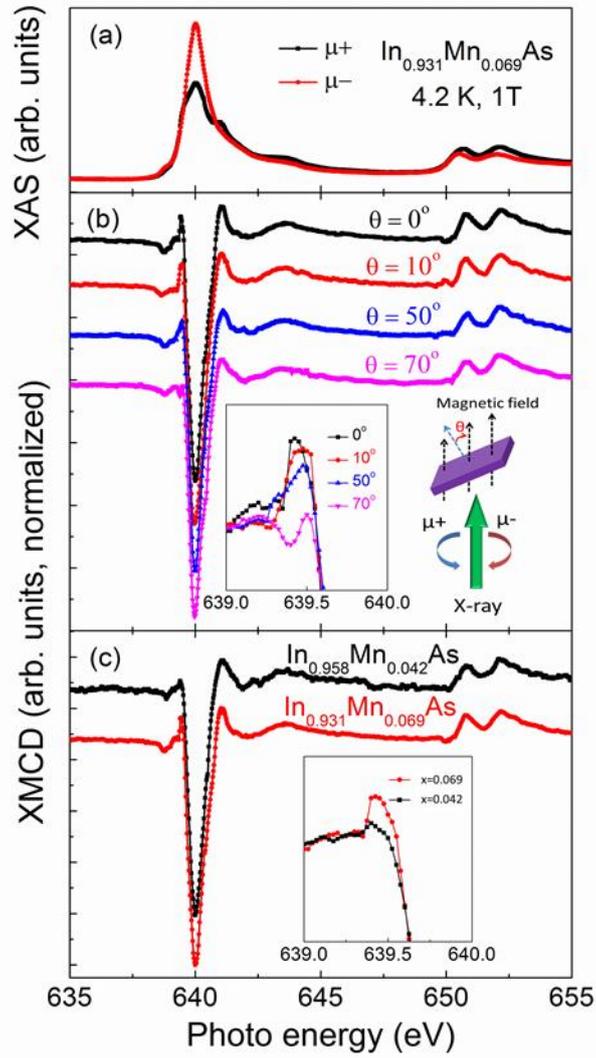

FIG. 5. (Color online) (a) XAS in parallel ($\mu^+$) and antiparallel ($\mu^-$) configurations for $In_{0.931}Mn_{0.069}As$ measured at 4.2 K under an external field of 1 T applied parallel to the layer surface-plane. (b) XMCD spectra: the sample was rotated from θ=0° (the magnetic field is applied perpendicular to the sample surface-plane, schematically shown as the right inset) to θ=70°. The inset shows the zoom at the Mn $L_3$ pre-edge at different angles. (c) XMCD spectra at θ=0° for $In_{1-x}Mn_xAs$ films with x=0.042 (square) and x=0.069 (circles). The inset shows the Mn $L_3$ pre-edge of InMnAs films with different Mn concentrations.